\documentclass[aps,prd,twocolumn,nofootinbib,longbibliography,10pt]{revtex4-2}
\usepackage{etoolbox}
\usepackage{dcolumn,tensor,nicefrac}
\usepackage{amsmath,amssymb,amsfonts,mathtools}
\usepackage{mathrsfs,bbold}
\usepackage{graphicx}
\usepackage[colorlinks=true,urlcolor=blue,citecolor=red,linkcolor=blue]{hyperref}
\usepackage{accents}
\newlength{\dhatheight}
\usepackage{natbib}
\usepackage{wrapfig}
\usepackage{flushend,BOONDOX-cal,BOONDOX-frak}

\makeatletter
\newsavebox{\@brx}
\newcommand{\llangle}[1][]{\savebox{\@brx}{\(\m@th{#1\langle}\)}%
  \mathopen{\copy\@brx\kern-0.5\wd\@brx\usebox{\@brx}}}
\newcommand{\rrangle}[1][]{\savebox{\@brx}{\(\m@th{#1\rangle}\)}%
  \mathclose{\copy\@brx\kern-0.5\wd\@brx\usebox{\@brx}}}
\makeatother
\begin{document}
\title{\textbf{Inverse logarithmic correction in the HBAR entropy of an atom falling into a renormalization group improved charged black hole}}
\author{Arpita Jana}
\email{janaarpita2001@gmail.com}
\affiliation{Department of Astrophysics and High Energy Physics, S. N. Bose National Centre for Basic Sciences, JD Block, Sector-III, Salt Lake City, Kolkata-700 106, India}
\author{Soham Sen}
\email{sensohomhary@gmail.com}
\affiliation{Department of Astrophysics and High Energy Physics, S. N. Bose National Centre for Basic Sciences, JD Block, Sector-III, Salt Lake City, Kolkata-700 106, India}
\author{Sunandan Gangopadhyay}
\email{sunandan.gangopadhyay@gmail.com}
\affiliation{Department of Astrophysics and High Energy Physics, S. N. Bose National Centre for Basic Sciences, JD Block, Sector-III, Salt Lake City, Kolkata-700 106, India}
\begin{abstract}
\noindent In this work, we have considered a spherically symmetric non-rotating charged black hole geometry where Newton's gravitational constant and the charge of the black hole flow with the energy scale. We have used the Kretschmann scale identification to write down the finite cutoff for the momentum scale regarding the proper distance. Introducing the flow of running couplings, the event horizon radius of the black hole using quantum-improved Reissner-Nordstrom metric was found in \href{https://doi.org/10.1103/PhysRevD.104.066016}{Phys. Rev. D 104 (2021) 066016}. We have, in this work, explored the thought experiment of a two-level atom freely falling into the event horizon of a quantum-improved charged black hole and have computed the transition probability of the atom for going from its ground state to the excited state via emission of a virtual photon. We find that the probability deviates slightly from the pure Planckian spectrum. We have shown that this deviation is due to the presence of an incomplete lower gamma function in the distribution function. We have then computed the horizon brightened acceleration radiation entropy and found that it is identical to the Bekenstein-Hawking entropy followed by the renormalization group correction terms including an inverse logarithmic and a square root of the area term due to emitting photons. 
\end{abstract}
\maketitle
\section{Introduction}
\noindent The historical development of modern theoretical physics started with the development of quantum mechanics, statistical physics, relativity, and other contemporary scientific developments at the beginning of the twentieth century. One of the most significant theoretical developments of that time was the theory of General relativity by Albert Einstein \cite{EinsteinGR1, EinsteinGR2} which is considered to be the most successful classical theory of gravity. It has been observed that among the four fundamental forces, electromagnetism, weak and strong nuclear forces have well-defined quantum field theoretic descriptions. In contrast, the unification of general relativity with quantum mechanics to unveil new horizons in the realm of gravity and its quantum nature is still not achieved. One of the more acceptable attempts to formulate this theory is the asymptotic safety scenario based on the renormalization group approach \cite{Reuter1,Reuter2,Percacci}. In 2000, Bonanno and Reuter first implicated this approach \cite{BonannoReuter} in Schwarzschild black hole geometry where the usual Newton's gravitational constant was replaced by running gravitational constant obtained from the renormalization group equation. After that, several researches on quantum-improved black holes and their properties including horizon structures, thermodynamic laws, and Hawking temperature have been executed in \cite{BonannoReuter2,ReuterTuiran,Falls,Koch1,Koch2,BonannoKoch,PawlowskiStock,Platania1,BonannoCasadio,Platania2,ReuterWeyer}.
\noindent Till this point, researchers have mostly worked on the quantum effects of the only gravitational field in the quantum-improved background, rather than the combined or individual quantum effects of other fundamental forces. The combined quantum effects of the electromagnetic field and gravity on spacetime structure in near singularity region were studied in \cite{HarstReuter,Christiansen, Eichhorn}.  In \cite{Ishibashi}, Ishibashi \textit{et. Al.} considered the quantum effects of electromagnetic field along with the gravitational field on the static spherically symmetric charged black hole geometry, where both the gravitational constant and charge of the black hole flow with the momentum cutoff scale. 
The usual metric for a spherically symmetric black hole can be written as \cite{BonannoReuter}
\begin{equation}\label{ds^2}
    ds^{2}=-f(r)dt^{2}+\frac{1}{f(r)}dr^{2}+r^{2}d\theta^{2}+r^{2}\sin^{2}{\theta}d\phi^{2}
\end{equation}
where $f(r)$ is called the lapse function of the black hole. For the classical Reissner-Nordstrom case, the form of the lapse function is
\begin{equation}\label{f0(r)}
    f(r)=1-\frac{2G_{0}M}{r}+\frac{G_{0}e^{2}_{0}}{r^{2}}
\end{equation}
where $G_{0}$ is Newton's gravitational constant and $e_{0}$ is the classical U(1) coupling.
So, for the quantum-corrected charged black hole case, the constants will be replaced by the running couplings and hence, the metric takes the form 
\begin{equation}\label{f(r)}
    f(r)=1-\frac{2G(r)M}{r}+\frac{G(r)e^{2}(r)}{r^{2}}
\end{equation}
where M is the mass of the black hole and $G(r)$ and $e(r)$ denote the running couplings obtained from the renormalization group flow equation \cite{BonannoReuter,Ishibashi}. In \cite{JanaSenGangopadhyay}, we have already considered the case of a quantum-corrected charged black hole where only Newton's gravitational constant flows with the momentum scale. The argument behind such an approximation lies in the fact that the running coupling becomes strong only near the Planck scale as has been discussed earlier in \cite{RuizTuiran}. Now, the study of acceleration radiation for a two-level atom falling into the event horizon of a black hole is quite important and initially was discussed in \cite{Scullypnas}. The most significant outcome is the fact that if one calculates the von Neumann entropy for a cloud of two-level atoms radially falling into the even horizon of the black hole then the rate of change of the entropy is the rate of change of the area of the black hole divided by four (in natural units with the Boltzmann constant set to unity). Hence, the entropy due to the radiation emitted from the atoms is similar to the usual Bekenstein-Hawking entropy \cite{Hawking1,Hawking2,Hawking3,Bekenstein1,Bekenstein2}. They named this entropy ``\textit{Horizon Brightened Acceleration Radiation entropy}" (HBAR entropy) to recognize it differently from Bekenstein-Hawking radiation. One of the main motivations of this work was to enlighten the unification of general relativity and quantum optics \cite{Weiss,Philbin}. So far, several analyses on acceleration-radiation and HBAR entropy for different black hole spacetimes have been done \cite{qsch,Camblong,kerr1,AziziCamblong1,AziziCamblong2,SenMandal,kerr2,DasSen}. In \cite{qsch}, it was theoretically shown that the HBAR entropy picks up a logarithmic correction where the black hole geometry was considered to be the renormalization group improved Schwarzschild black hole. This type of logarithmic correction in the Bekenstein-Hawking entropy was first proposed in \cite{KaulMajumdar} and was claimed to be universal in nature for a quantum gravity setting. Later in \cite{JanaSenGangopadhyay}, the same structure of the HBAR entropy with subleading logarithmic correction was obtained and was claimed to be universal in nature.

\noindent In this work, we extend the work presented in \cite{JanaSenGangopadhyay} by considering the quantum effects of the electromagnetic field. Hence, one needs to deal with Landau poles involving the divergence of running coupling at a finite momentum scale which has been discussed in detail in \cite{Ishibashi}. Hence, one needs to find a proper scale identification \cite{BonannoEichhorn} in terms of the radial distance $r$. Several schemes for this scale identification have been brought forward so far. But, as discussed in \cite{PawlowskiStock, Ishibashi}, we need to use the Kretschmann scalar to fix the momentum cutoff scale. The main advantage of choosing this is that it is a diffeomorphism invariant scheme and it can easily compare the Schwarzschild and the Reissner-Nordstrom solution. Our main aim in this work is to consider the flow of the charge of the black hole along with the flow of Newton's gravitational constant and observe the importance of the running couplings corresponding to the charge parameter in the HBAR entropy. We are also interested in seeing whether the probability distribution will follow the standard Planckian distribution observed in \cite{Scullypnas,qsch,JanaSenGangopadhyay} which will be used along with the absorption probability to calculate the von Neumann entropy for the system.

\noindent This paper is organized as follows. In section \ref{S2}, we provide a brief review of a quantum-corrected charged black hole spacetime where the exact renormalization group flow equations corresponding to Newton's gravitational constant and the charge are discussed. We discuss the Kretschmann scale identification in short and write down the event horizon for this black hole. Then in section \ref{S3}, we calculate the atomic trajectories in this black hole spacetime, and then we move to calculate the transition probability of the atom going from its ground state to the first excited state when a single virtual photon is emitted simultaneously. In section \ref{S4}, we compute the HBAR entropy and finally in section \ref{S5}, we summarize and conclude our whole analysis.

\section{Renormalization group improved black hole spacetime}\label{S2}
\subsection{RG flow of couplings}
\noindent The first introduction of U(1) gauge theory was to describe quantum electrodynamics in flat spacetime which was later implemented in curved spacetime. To consider the quantum effects of the electromagnetic field in the blackhole spacetime, the authors in \cite{Eichhorn} implement a method in which by using the functional renormalization group in U(1) coupled gravity theory one achieves the UV completion of U(1) gauge theory. Following their path, an extensive breakdown of quantum-improved charged black hole was made in \cite{Ishibashi} with the consideration of running Newton's constant and U(1) gauge coupling. The standard scheme to write down a quantum-improved geometry, as has been discussed in \cite{Ishibashi}, can be summarized into a few crucial steps.
\begin{enumerate}
\item At first, one needs to consider the exact renormalization group equations and from there obtain the analytical form of the scale-dependent coupling constants.
\item The next step is to identify the cutoff scale $k=k(r)$ such that the running couplings can be expressed as a function of the physically significant parameter in consideration. This cutoff scale identification is mostly done such that $k\propto 1/\chi(r)$ with $\chi(r)$ being a function of the radial distance. Another way is to identify the cutoff scale with respect to geometric curvature scalar quantities like the Ricci scalar $R$, $R_{\alpha\beta}R^{\alpha\beta}$, and the Kretschmann scalar $R_K=R_{\alpha\beta\gamma\delta}R^{\alpha\beta\gamma\delta}$.
\end{enumerate}
One can, therefore, simply start by writing down the exact renormalization group equations corresponding to Newton's gravitational constant and the charge as \cite{HarstReuter, Eichhorn}
\begin{align}
k\frac{d\tilde{G}(k)}{dk}&=2\tilde{G}(k)\left(1-\frac{\tilde{G}(k)}{4\pi\tilde{\alpha}}\right)\label{flow1}\\
k\frac{de(k)}{dk}&=\frac{e(k)}{4\pi}\left(\frac{be^2(k)}{4\pi^2}-\tilde{G}(k)\right)\label{flow2}
\end{align}
where $\tilde{\alpha}$ and b are numerical parameters correspond to the fixed points $\tilde{G}_{*}$ and $e^{2}_{*}$ as
\begin{equation}\label{fixedpoint}
    \tilde{G}_{*} = 4\pi \tilde{\alpha} , ~e^{2}_{*} = (4\pi)^{2}\frac{\tilde{\alpha}}{b}~.
\end{equation}
In eq.(s)(\ref{flow1},\ref{flow2})  $\tilde{G}(k)$ gives the diensionless Newton's constant which is defined as $\tilde{G}(k)\equiv k^{2}G(k)$. Solving the exact renormalization group flow equation corresponding to Newton's gravitational constant in eq.(\ref{flow1}), one obtains the analytical form of $G(k)$ as a function of the momentum $k$ as \cite{BonannoReuter}
 \begin{equation}\label{Gflow}
     G(k) = \frac{G_{0}}{1+\frac{G_{0}k^{2}}{4\pi \tilde{\alpha}}}
 \end{equation}
where $G_{0}= G(k=0)$ is the value of Newton's gravitational constant at the current time. The allowed values of $\tilde{\alpha}$ and $b$ are explicitly discussed in \cite{Ishibashi}. Following the arguments given in \cite{Ishibashi}, we also took the range of $\tilde{\alpha}$ to be $0< \tilde{\alpha}\leq 1$. In our analysis, we shall mainly focus on the general case $\tilde{\alpha}\ne 1$ i.e. $0<\tilde{\alpha}<1$.

\noindent In order to find the flow of the charge, one now needs to solve the flow equation corresponding to the charge parameter given in eq.(\ref{flow2}). As discussed in \cite{Ishibashi}, one can define two new functions $p(k)\equiv\frac{\tilde{\alpha}G_{0}k}{4\pi \tilde{\alpha}+G_{0}k^{2}}$ and $q(k)\equiv\frac{b}{(4\pi)^{2}k}$ in terms of which one can  rewrite the flow equation eq.(\ref{flow2}) in similar form to that of the Bernoulli equation as
\begin{equation}\label{Bernoulli}
    \frac{de(k)}{dk}+p(k)e(k) = q(k)e^{3}(k)~.
\end{equation}
One can now solve the above equation and obtain the analytical form of the flow of charge as a function of $k$ to be
\begin{equation}\label{flow}
\begin{split}
    \frac{1}{e^{2}(k)} &= C_{0}(1+\mathfrak{D}k^{2})^{\tilde{\alpha}}\\& + \frac{\tilde{\alpha}(1+\mathfrak{D}k^{2})}{e^{2}_{*}(1-\tilde{\alpha})}\tensor[_2]{F}{_1}\left(1,1-\tilde{\alpha},2-\tilde{\alpha},1+\mathfrak{D}k^{2}\right)
    \end{split}
\end{equation}
where $\tensor[_2]{F}{_1}$ denotes the Gauss hypergeometric function, $C_{0}$ denotes an integration constant, and the constant $\mathfrak{D}$ is defined as $\mathfrak{D} \equiv \frac{G_{0}}{4\pi \tilde{\alpha}}$. Using the properties of the hypergeometric functions \cite{hypergeometric}, one can recast the above equation as \cite{Ishibashi}
\begin{equation}\label{flow_simplified}
\begin{split}
    &\frac{1}{e^{2}(k)} = C_{0}(1+\mathfrak{D}k^{2})^{\tilde{\alpha}} + \frac{\tilde{\alpha}}{e^{2}_{*}}\sum \limits_{n = 0}^{\infty}\frac{\Gamma(\tilde{\alpha}+n)}{(n!)^{2}\Gamma(n)}\\ \times&\left[\psi(n+1)-\psi(\tilde{\alpha}+n)-\log\left[\frac{\mathfrak{D}k^{2}}{1+\mathfrak{D}k^{2}}\right]\right]\left[\frac{\mathfrak{D}k^{2}}{1+\mathfrak{D}k^{2}}\right]^{n}
    \end{split}
\end{equation}
with $\psi$ denoting the digamma function. As one goes away from the singularity, it is expected that we approach the infrared limit. Hence for very small $k$ limit, eq.(\ref{flow_simplified}) can be approximately written as
\begin{equation}\label{eflow}
    \frac{1}{e^{2}(k)} \simeq C_{0} - \frac{\tilde{\alpha}}{e^{2}_{*}} (\gamma + \psi(\tilde{\alpha}))+\frac{\tilde{\alpha}}{e^{2}_{*}}\log\left[\frac{1+\mathfrak{D}k^{2}}{\mathfrak{D}k^{2}}\right]
\end{equation}
where the Euler constant is given by $\gamma = -\psi(1)$. In this equation, the second term is finite for small values of the constant $\tilde{\alpha}$ as $\tilde{\alpha}(\gamma + \psi(\tilde{\alpha})) = -1 +O(\tilde{\alpha}^{2})$. The running of the coupling depends on the value of the constant $C_0$ and as has been discussed in detail in \cite{Ishibashi}, it can be separated into three parts:
\begin{enumerate}
    \item For $C_{0}>0$, the U(1) coupling exists and it is valid in the rangle $k\in(0,\infty)$. As $k\rightarrow\infty$, the running coupling vanishes which gives the asymptotic limit of the coupling constant for the case in consideration.
    \item For $C_{0}<0$, the U(1) coupling diverges and a Landau pole comes into considerations. The coupling stops at such a pole.
    \item Finally, for $C_{0}=0$, the running coupling becomes constant asymptotically.
\end{enumerate}
As discussed in \cite{Ishibashi}, we will use $C_{0}=0$ throughout our analysis.

\noindent With the forms of Newton's gravitational constant and the charge of the black hole in hand, the next step is to determine the scale identification for $k$ in terms of the radial distance $r$. For a spherically symmetric black hole spacetime, the momentum cutoff scale $k$ in general is identified with the radial distance $r$ as \cite{BonannoReuter,Ishibashi}
\begin{equation}
    k = \frac{\eta}{d(r)}
\end{equation}
where $\eta$ is a constant and $d(r)$ denotes some function of $r$ with inverse momentum dimension in natural units. Several frameworks have been recommended to fix the scale identification so far. In the current scale, the authors in \cite{Ishibashi} have used the Kretschmann scalar defined as $R_K\equiv R_{\alpha \beta \gamma \delta}R^{\alpha \beta \gamma \delta}$ to identify the momentum cutoff scale. It was first proposed in \cite{PawlowskiStock} as the Kretschmann scalar is a diffeomorphism invariant quantity that has a dimension equal to the fourth power of momentum in natural units. Hence, the scale identification in terms of the Kretschmann scalar is given by \cite{Ishibashi}
\begin{equation}\label{Kdef}
    k^{4} = \chi^{4}R_K(r)
\end{equation}
with $\chi$ being anumber\footnote{For a detailed discussion on the possible values of the constant $\chi$ please see \cite{PawlowskiStock}}. The analytical form of the Kretschmann scalar for a classical Reissner-Nordstr\"{o}m metric is obtained as \cite{Ishibashi}
\begin{equation}\label{Kretschmann}
    R_K(r)=\frac{8G_{0}^{2}}{r^{8}}(6M^{2}r^{2}-12Me_{0}^{2}r+7e_{0}^{4})
\end{equation}
where $e_{0}$ can be obtained from the following equation \cite{Ishibashi}
\begin{equation}\label{e0}
\frac{1}{e_0^2}=C_0(1+\mathfrak{D}k_{L}^2)^{\tilde{\alpha}}+\frac{1}{e_*^2}\tensor[_2]{F}{_1}\left[1,\tilde{\alpha},1+\tilde{\alpha},\frac{1}{1+\mathfrak{D}k_L^2}\right]
\end{equation}
with $k_L\equiv 1$ GeV. Substituting the form of the Kretschmann scalar from eq.(\ref{Kretschmann}) in eq.(\ref{Kdef}), one obtains the cutoff scale identification to be
\begin{equation}\label{Identification}
k^4(r)=\frac{8G_0^2\chi^4}{r^8}\left(6M^2r^2-12e_0^2Mr+7e_0^4\right)~.
\end{equation}
 As one approaches the high energy limit (i.e. $k\rightarrow\infty$) then it is equivalent to approaching the $r=0$ point and as a result, the above scale identification takes the form
\begin{equation}\label{high_energy_identification}
k^{2}(r)\simeq \frac{2\sqrt{14}\chi^{2}G_{0}e_{0}^{2}}{r^{4}}~.
\end{equation}
Now, the case $k\ll 0$ is equivalent to going far away from the $r=0$ point of the black hole, and as a result in eq.(\ref{Identification}), the first term inside of the parentheses in the right-hand side of the equation dominates. Hence, one can recast eq.(\ref{Identification}) as
\begin{align}\label{low_energy_identification}
    k^{2}(r)\simeq \frac{4\sqrt{3}\chi^{2}G_{0}M}{r^{3}}~.
\end{align}
Substituting the above scale identification for $r\gg 0$ in eq.(s)(\ref{Gflow},\ref{eflow}), we can write down the analytical form of $G(r)$ and $e(r)$ as
\begin{equation}\label{Couplings_in_r}
\begin{split}
    G(r)&\simeq G_{0}\left(1-\frac{\sqrt{3}\chi^2 G_0^2M}{\pi\tilde{\alpha}r^3}\right)\\e^{2}(r) &\simeq \frac{e^{2}_{*}}{\tilde{\alpha}\log\left[\frac{1}{\mathfrak{D}k^{2}(r)}\right]} = \frac{e^{2}_{*}}{\tilde{\alpha}\log\left[\frac{\pi \tilde{\alpha}r^{3}}{\sqrt{3}\chi^{2}G_{0}^{2}M}\right]}~
\end{split}
\end{equation}
where to obtain $e(r)^2$, the constant $C_0$ is set to zero and $\frac{1}{e_*^2}$ is neglected with respect to the third term in eq.(\ref{eflow}), as for small $k$ values this logarithmic term is larger. In the next subsection, the form of $G(r)$ and $e^2(r)$ will be used to write down the analytical form of the lapse function $f(r)$ that we shall use to consider the acceleration-radiation phenomenon.
\subsection{Metric and the horizon of the black hole}
\noindent For a renormalization group improved black hole spacetime where both $G$ and $e$ flow with the momentum scale, the lapse function is given by eq.(\ref{f(r)}). Substituting the forms of $G(r)$ and $e(r)$ from eq.(\ref{Couplings_in_r})\cite{RuizTuiran,Ishibashi} one can express $f(r)$ as
\begin{equation}\label{f(r)_RG_improved}
\begin{split}
    f(r)&= 1-\frac{2G(r)M}{r}+\frac{G(r)e^{2}(r)}{r^{2}}\\
    &\simeq1-\frac{2G_0M}{r}+\frac{e_*^2G_0}{r^2}\frac{1}{\tilde{\alpha}\ln\left[\frac{\pi\tilde{\alpha}r^3}{\sqrt{3}\chi^2G_0^2M}\right]}\\&+\frac{2\sqrt{3}\chi^2G_0^4M^2}{\pi\tilde{\alpha}r^4}-\frac{\sqrt{3}\chi^2G_0^2M}{\pi\tilde{\alpha}r^5}\frac{e_*^2 G_0}{\tilde{\alpha}\ln\left[\frac{\pi\tilde{\alpha}r^3}{\sqrt{3}\chi^2G_0^2M}\right]}~.
\end{split}
\end{equation}
As we shall be working in the vicinity of the event horizon of the black hole, for a general black hole that has mass equivalent to or more than the solar mass $r\gg 0$. As a result, one can neglect the last two terms on the right-hand side of the above equation and recast it as
\begin{equation}\label{f(r)_Simplified}
    f(r)\simeq 1-\frac{2G_{0}M}{r}+\frac{G_{0}e_{*}^{2}}{r^{2}}\frac{1}{\tilde{\alpha}\ln\left[\frac{\pi\tilde{\alpha}r^{3}}{\sqrt{3}\chi^{2}G_{0}^{2}M}\right]}~.
\end{equation}
Another important reason for the above approximation lies in the fact that we want to primarily investigate the effect of the charge renormalization on the HBAR entropy as we have already investigated the effect of the flow of $G(r)$ in \cite{JanaSenGangopadhyay}. To easily write down the form of the lapse function and for the sake of analytical simplicity, we shall put $2G_{0}M = 1$ (which shall be restored later while calculating the HBAR entropy) and define two constants as $\epsilon\equiv\frac{e_{*}^{2}}{2M\tilde{\alpha}}$, and $\xi\equiv \frac{4\pi M\tilde{\alpha}}{\sqrt{3}\chi^{2}}$. One can then express the analytical form of $f(r)$ from eq.(\ref{f(r)_Simplified}) in a simplified form\footnote{In the current analysis $\log[x]=\log_e[x]=\ln[x]$.}
\begin{equation}\label{lapsefunc}
    f(r)= 1-\frac{1}{r}+\frac{\epsilon}{r^{2}\ln\left[\xi r^{3}\right]}~.
\end{equation}
\noindent Since we aim to find the trajectories of the freely falling atom in the near horizon region, we need to determine the analytical form of the event horizon. For the current analysis, the event horizon $r_{+}$ can be written in terms of the Schwarzschild radius $r_{s}$ as $r_{+} = r_{s}+\epsilon \delta r$, where $r_{s}$ is given by $r_{s}=\frac{2G_{0}M}{c^{2}}$ and the small $\delta r$ correction includes the contribution from the charge renormalization. Here $r_s=1$ as $2G_0M=1$ and $c=1$ as well which shall be restored later. From the definition of $e_{*}^{2}$ in eq.(\ref{fixedpoint}) and the values of $b$ discussed in \cite{Ishibashi}, we can safely assume that the term $\frac{e_{*}^{2}}{\tilde{\alpha}}$ is very small and as a result $\epsilon$ can be considered to be a small quantity as well. On the event horizon, the lapse function $f(r)$ vanishes and it can be expressed as $f(r_{+})=0$. Hence, we obtain the analytical form of the correction term to be $\delta r = -\frac{1}{\log {\xi}}$ and as a result the event horizon radius reads 
\begin{equation}\label{horizon}
    r_{+} =r_s+\epsilon\delta r\simeq 1-\frac{\epsilon}{\log{\xi}}~.
\end{equation}

\section{Excitation probability of atom falling into the quantum-improved charged black hole}\label{S3}
\noindent In this section, we shall consider the thought experiment discussed in \cite{Scullypnas}, where a two-level atom was considered to fall into the event horizon of a Schwarzschild black hole. In our analysis of renormalization group improved charged black hole, we also stick to this conjecture with the consideration of running couplings. The atom with angular frequency $\Omega$ is falling along the radial trajectory from infinity with zero initial velocity. The atom trajectories can be written in terms of the lapse function $f(r)$ of the black hole as 
\begin{equation}\label{trajeq}
\tau(r)=-\int\frac{dr}{\sqrt{1-f(r)}}~,~~t(r)=-\int \frac{dr}{f(r)\sqrt{1-f(r)}}~.
\end{equation}
where $\tau$ denotes the proper time for the atom. We now do a change of variable $z = r - r_{+}$ so that at the event horizon $z$ becomes zero. As we are very near the event horizon of the black hole, $r-r_+\ll 1$, which lets us consider $z$ as a very small quantity. Using the form of $f(r)$ from eq.(\ref{lapsefunc}), we can recast the trajectory equations in eq.(\ref{trajeq}) as a function of $z$ as
\begin{align}\label{trajectories}
    \tau(z) \simeq &-\int dz\left(1+\frac{z}{2}-\frac{3\epsilon z}{2(\log{\xi})^{2}}\right) \nonumber\\
    \simeq &-z-\frac{z^{2}}{4}\left[1-\frac{3\epsilon}{(\log {\xi})^{2}}\right]+C_{1} \\
    t(z) \simeq & -\int
    \frac{dz}{z}\left(1+\frac{z}{2}-\frac{3\epsilon z}{2(\log\xi)^2}\right)\left(1+\frac{3\epsilon}{(\log\xi)^2}+z\right.\nonumber\\&\left.-\frac{3\epsilon z}{2(\log\xi)^2}-\frac{9\epsilon z}{(\log \xi)^3}\right)\nonumber\\ \simeq&- \left[1+\frac{3\epsilon}{(\ln {\xi})^{2}}\right]\ln{z}-\left[\frac{3}{2}-\frac{3\epsilon}{2(\ln{\xi})^{2}}-\frac{9\epsilon}{(\ln{\xi})^{3}}\right]z\nonumber\\&+C_{2}
\end{align}
where $C_{1}$ and $C_{2}$ are integration constants. For the expression of $\tau(z)$, up to order $z^{2}$ is taken as when we shall derive the excitation probability, then we need to get the analytical form of the function $\frac{d\tau}{dz}$ up to first order in $z$. The field emitted due to acceleration radiation can be safely approximated by a scalar field \cite{Scullypnas}. The covariant Klein-Gordon equation for a scalar field\footnote{It is important to note that as we are more interested in the coordinate dependence of the field, the implementation of a vector field for the photons will not provide any significant alterations to the transition probabilities of the two-level atom while emitting and absorbing photons.} $\psi(t,\vec{x})$ has the form
\begin{equation}\label{KG}
\begin{split}
\frac{1}{\sqrt{-g}}\partial_\mu\left(\sqrt{-g}g^{\mu\nu}\partial_\nu\right)\psi(t,\vec{x})-m^2\psi(t,\vec{x})=0~.
\end{split}
\end{equation}
As we are considering scalar photons, it is safe to consider $m$ to be zero in the above equation. Now, neglecting the angular momentum part we can reduce the covariant massless Klein-Gordon equation into an effective (1+1)-dimensional case where the surviving coordinates are just the time and radial distance. Making use of the separation of variables technique as $\psi (t,r) = T(t)R(r)$, one can recast eq.(\ref{KG}) as (for $m=0$)
\begin{equation}
\frac{1}{T(t)}\frac{d^{2}T(t)}{dt^{2}}-\frac{f(r)}{r^{2}R(r)}\frac{d}{dr}\left(r^{2}f(r)\frac{dR(r)}{dr}\right) = 0~.
\end{equation}
The general solution of the above equation reads
\begin{equation}\label{wavefunc}
    \psi(t,r) = \frac{\Psi_{\nu}(t,r)}{r}
\end{equation}
where  $\Psi_{\nu}(t,r)$ is given by
\begin{equation}\label{wave}
    \Psi_{\nu}(t,r) = \exp\left[i\nu(t(r)-r_{*}(r))\right]
\end{equation}
with $\nu$ being the field frequency. In the above equation $r_{*}(r)$ is the Regge-Wheeler coordinate for a scalar photon which is defined by $r_{*}(r)=\int \frac{dr}{f(r)}$. The analytical form of $r_*$ as function of $z$ is
\begin{equation}\label{rwcoord}
    r_{*}(z) \simeq \left[1+\frac{3\epsilon}{(\ln{\xi})^{2}}\right]\ln{z} + \left[1-\frac{3\epsilon}{2(\ln{\xi})^{2}}-\frac{9\epsilon}{(\ln{\xi})^{3}}\right]z + C_{3}
\end{equation}
where $C_{3}$ is an integration constant. Now, using eq.(s)(\ref{trajectories},\ref{wavefunc},\ref{wave},\ref{rwcoord}), one can obtain finally 
\begin{equation}\label{finalwavefunc}
\begin{split}
    \psi(t,z) =&\frac{\exp\left[i\nu(t(r)-r_{*}(r))\right]}{r}\\ \simeq&\left(1-z+\frac{\epsilon}{\ln{\xi}}(1-2z)\right)z^{-2i\nu\left(1+\frac{3\epsilon}{(\ln{\xi})^{2}}\right)}\\
    &\times \exp\left[-i\nu z\left(\frac{5}{2}+\frac{3\epsilon}{(\ln\xi)^2}\left(1+\frac{6}{\ln\xi}\right)\right)\right]~.
    \end{split}
\end{equation}
The above form of the scalar field solution is quite important as we have considered the $\frac{1}{r}$ contribution and progressed beyond the plane wave approximation. Initially in all of the previous similar analyses  \cite{Scullypnas,qsch,Camblong,kerr1,AziziCamblong1,AziziCamblong2,SenMandal,kerr2,DasSen}, the solution considered is $\exp\left[i\nu(t(r)-r_{*}(r))\right]$ by ignoring the $\frac{1}{r}$ contribution which made the solution of the scalar field to be plane wave-like. The reason behind such approximation lies in the fact that the $\frac{1}{r}$ contribution can be expressed as $\frac{1}{1+\delta}$ $(\delta\ll 1)$ for a near horizon analysis which can be neglected eventually. The form of the scalar field taken in our current work is more logically sound and a step towards a more realistic scenario.

\begin{figure}
\begin{center}
\includegraphics[scale=0.28]{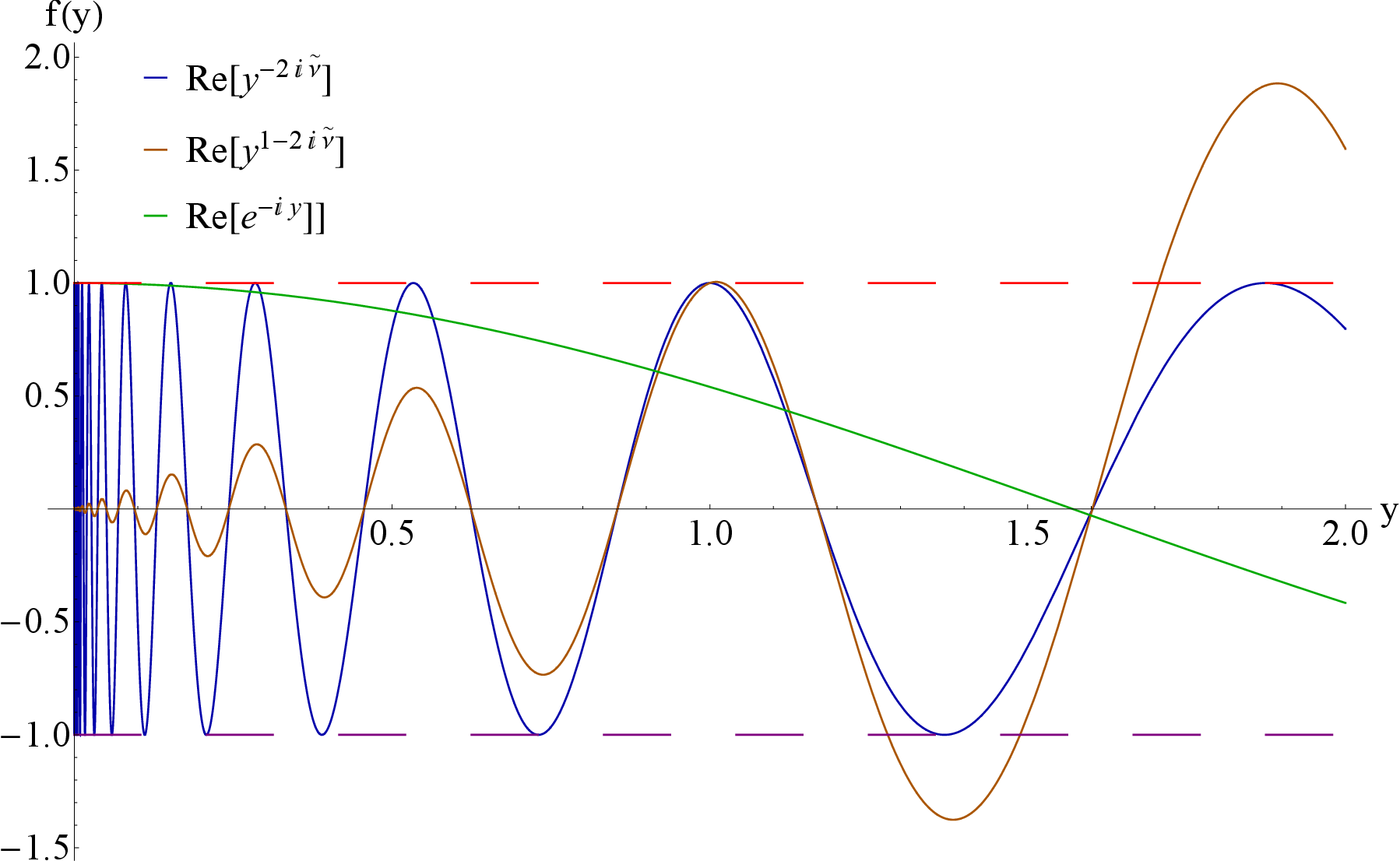}
\caption{Plot of the functions $f(y)$ against $y$ where $f(y)=\Re[y^{-2i\tilde{\nu}}],~\Re[y^{1-2i\tilde{\nu}}]$, and $\Re[\exp[-iy]]$ where $\tilde{\nu}=5$.\label{P1}}
\end{center}
\end{figure}

\begin{figure}
\begin{center}
\includegraphics[scale=0.28]{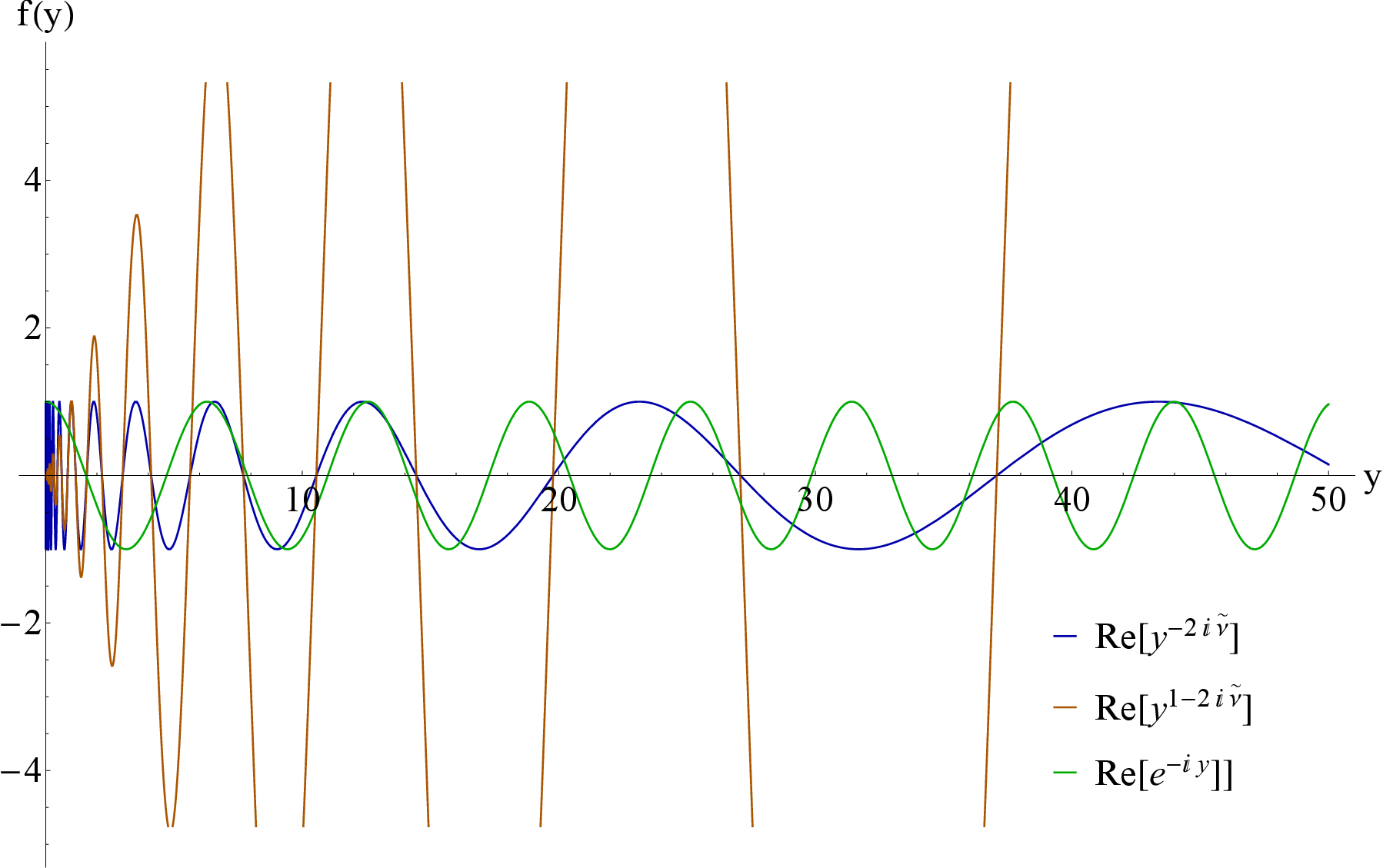}
\caption{Plot of the functions $f(y)$ against $y$ where $f(y)=\Re[y^{-2i\tilde{\nu}}],~\Re[y^{1-2i\tilde{\nu}}]$, and $\Re[\exp[-iy]]$ where $\tilde{\nu}=5$ depicting the large $y$ behaviour of the functions.\label{P3}}
\end{center}
\end{figure}

\noindent We shall now move towards our main goal of finding the excitation probability of the two-level atom (freely falling into the event horizon of a quantum-corrected charged black hole geometry) to go from its ground state to excited state with the simultaneous emission of a virtual photon. As discussed in \cite{Scullypnas}, a mirror can be placed on the event horizon of the black hole to shield the Hawking radiation from interacting with the field and the atom. The atom-field interaction Hamiltonian is given as
\begin{equation}\label{inthamiltonian}
\hat{V}_I(\tau)=\hbar \mathcal{G}[\hat{b}_\nu \psi_\nu(t(\tau),r(\tau))+H.c.][\hat{\zeta}e^{-i\Omega\tau}+h.c.]
\end{equation}
where $\hat{b}_{\nu}$ is the annihilation operator corresponding to the scalar field, $\mathcal{G}$ denotes the atom-field coupling and $\hat{\zeta} = |g\rangle\langle e|$ where $|g\rangle$ and $|e\rangle$ give the ground and excited states of the atom respectively, and $\Omega$ denotes the transition frequency of the atom. Initially, the atom is in the ground state and there is no scalar photon where the field is initially considered to be in the Boulware vacuum \cite{Boulware}. The initial state of the system then reads $|i\rangle\equiv|0_\nu,g\rangle\equiv |0_\nu\rangle\otimes |g\rangle$. The atom then reaches its excited state with the simultaneous emission of a scalar photon and as a result, the final state of the system reads $|f\rangle\equiv|1_\nu,e\rangle$. Hence, the excitation probability of the system going from $|i\rangle$ to $|f\rangle$ can be expressed as
\begin{align}\label{Prob}
    \mathcal{P}_{\text{exc}} = &\frac{1}{\hbar^{2}}\left|\int d\tau \langle f|\hat{V}_{I}(\tau)|i\rangle d\tau \right|^{2} \nonumber\\
    =& \mathcal{G}^{2}\left|\int dz \left(\frac{\partial \tau}{\partial z}\right)\frac{\Psi_\nu(t,z)}{r(z)}e^{i\Omega\tau(z)}\right|^{2}\nonumber \\
    \simeq & \mathcal{G}^{2}\Bigr|\int dz \left[1-\frac{z}{2}+\frac{\epsilon}{\ln{\xi}}\left(1-\frac{3z}{2\ln{\xi}}-\frac{3z}{2}\right)\right]\nonumber\\\times &z^{-2i\nu\left(1+\frac{3\epsilon}{(\ln{\xi})^{2}}\right)}e^{-i\Omega z \left[1+\frac{\nu}{\Omega}\left(\frac{5}{2}-\frac{3\epsilon}{(\ln{\xi})^{2}}\left(1+\frac{6}{\ln{\xi}}\right)\right)\right]}\Bigr|^{2}.
\end{align}
We shall now make a change of variables given as
\begin{equation}\label{y}
    y = z\Omega \left( 1+\frac{\nu}{\Omega}\left(\frac{5}{2}-\frac{3\epsilon}{(\ln{\xi})^{2}}\left(1+\frac{6}{\log{\xi}}\right)\right)\right).
\end{equation}
\medskip
\noindent Now, using the redefined variable from the above equation and substituting it into eq.(\ref{Prob}), it is possible to write down the analytical form of the transition probability as
\begin{widetext} 
\begin{equation}
    \begin{split}
        \mathcal{P}_{\text{exc}} \simeq &\frac{\mathcal{G}^{2}}{\Omega^{2}}\left[1-\frac{2\nu}{\Omega}\left(\frac{5}{2}-\frac{3\epsilon}{(\ln{\xi})^{2}}\left(1+\frac{6}{\log{\xi}}\right)\right)\right] \biggr|\int_{0}^{y_{f}} dy \biggr[\left(1+\frac{\epsilon}{\ln{\xi}}\right)- \frac{y}{2\Omega}\left[1+\frac{3\epsilon}{\log{\xi}}\left(1+\frac{1}{\log{\xi}}\right)\right.\\&\left.-\frac{\nu}{\Omega}\left(\frac{5}{2}-\frac{3\epsilon}{(\ln{\xi})^{2}}\left(1+\frac{6}{\log{\xi}}\right)+\frac{15\epsilon}{2\log{\xi}}\left(1+\frac{1}{\log{\xi}}\right)\right)\right]\biggr] y^{-2i\nu\left(1+\frac{3\epsilon}{(\ln{\xi})^{2}}\right)} e^{-iy}\biggr|^{2}   
        \end{split}
\end{equation}
where $y_{f}$ denotes the final limit of integration. This equation can also be written in a more compact form
\begin{equation}\label{totalprob}
    \mathcal{P}_{\text{exc}} \simeq \frac{ \mathcal{A}_\nu\mathcal{G}^{2}}{\Omega^{2}} \left|\int_{0}^{y_{f}} dy \left[\left(1+\frac{\epsilon}{\ln{\xi}}\right)-\frac{\mathcal{B}_\nu y}{2\Omega}\right]y^{-2i\nu\left(1+\frac{3\epsilon}{(\ln{\xi})^{2}}\right)} \exp[-iy]\right|^{2} 
\end{equation}
with $\mathcal{A}$ and $\mathcal{B}$ being defined as
\begin{equation}\label{Redefinition}
   \begin{split}
    \mathcal{A}_\nu\equiv&1-\frac{2\nu}{\Omega}\left(\frac{5}{2}-\frac{3\epsilon}{(\ln{\xi})^{2}}\left(1+\frac{6}{\log{\xi}}\right)\right) \\
    \mathcal{B}_\nu\equiv&1+\frac{3\epsilon}{\log{\xi}}\left(1+\frac{1}{\log{\xi}}\right)-\frac{\nu}{\Omega}\left(\frac{5}{2}-\frac{3\epsilon}{(\ln{\xi})^{2}}\left(1+\frac{6}{\log{\xi}}\right)+\frac{15\epsilon}{2\log{\xi}}\left(1+\frac{1}{\log{\xi}}\right)\right)~.
    \end{split}
\end{equation}
\end{widetext}
One can now write down a modified frequency $\tilde{\nu}=\nu \left(1+\frac{3\epsilon}{(\ln\xi)^2}\right)$, in terms of which we can recast the probability in eq.(\ref{totalprob}) as
\begin{equation}\label{Probability_redefinition}
\begin{split}
\mathcal{P}_{\text{exc}}=\frac{\mathcal{A}_\nu\mathcal{G}^2}{\Omega^2}\biggr|\left(1+\frac{\epsilon}{\ln\xi}\right)\mathcal{I}^{\tilde{\nu}}_1-\frac{\mathcal{B}_\nu}{2\Omega}\mathcal{I}^{\tilde{\nu}}_2\biggr|^2
\end{split}
\end{equation}
where the integrals $\mathcal{I}^{\tilde{\nu}}_1$ and $\mathcal{I}^{\tilde{\nu}}_2$ are defined as
\begin{align}
    \mathcal{I}^{\tilde{\nu}}_1&\equiv\int_{0}^{y_{f}} dy\exp[-iy]~y^{-2i \tilde{\nu}}\label{I1}\\
   \mathcal{I}^{\tilde{\nu}}_2 &\equiv \int_{0}^{y_{f}} dy\exp[-iy]~y^{1-2i\tilde{\nu}} ~.\label{I2}
\end{align}
Hence, obtaining the form of the transition probability in eq.(\ref{Probability_redefinition}) boils down to the task of solving the above two integrals. In eq.(\ref{I1}), there are two oscillatory functions $y^{-2i\tilde{\nu}}$ and $e^{-iy}$. For higher values of $y_{f}$, $e^{-iy}$ behaves as a highly oscillating function, while the $y^{-2i\bar{\nu}}$ function shows a ``Russian doll like behaviour" \cite{kerr1,kerr2} and hence, $y^{-2i\bar{\nu}}$ becomes slowly varying compared to $e^{-iy}$ for higher values of the uper limit of integration. We can depict this behaviour by plotting the real values of the functions $\exp[-iy]$, $y^{-2i\tilde{\nu}}$, and $y^{1-2i\tilde{\nu}}$ with respect to the change in $y$ for a fixed value of the dimensionless frequency $\tilde{\nu}$ in Fig.(\ref{P1}). It is important to note that it is more prudent to take the value of the dimensionless frequency to be less than unity but to truly depict the functional behaviour we have set $\tilde{\nu}=5$ in Fig.(\ref{P1}).
From Fig.(\ref{P1}), we observe that $\exp[-iy]$ varies very slowly in the $y\rightarrow 0$ regime to $y=1$ regime with the amplitude of oscillation being contained in the $[-1,1]$ range. For the $y^{1-2i\tilde{\nu}}$ function, we can see that for $y>1$, the amplitude of the function keeps on increasing beyond the $[-1,1]$ range. Hence, it is evident that the amplitude of the joint function $\exp[-iy]~y^{1-2i\tilde{\nu}}$ will keep on increasing which restricts one from increasing the upper limit of integration to infinity in eq.(\ref{I2}). It is also important to note that $y_f<1$ in eq.(\ref{I2}) as we are in the near horizon regime. Now, for the first integral in eq.(\ref{I1}), we can see from Fig.(\ref{P1}) that the $y^{-2i\tilde{\nu}}$ is highly oscillating in the near horizon regime with respect to the $\exp[-iy]$ function contributing to dominant contribution to the integral. To depict the long-term behaviour of the functions plotted in Fig.(\ref{P1}), we need to plot for higher values of $y$ which is executed in Fig.(\ref{P3}).
From Fig.(\ref{P3}), we observe that for high values of $y$, $y^{-i\tilde{\nu}}$ becomes slowly varying corresponding to the function $\exp[-iy]$. Hence, the primary contribution to the integral will come from small $y$ values which helps us to increase the upper limit of integration to $\infty$ in Fig.(\ref{I1}).
Hence, the upper limit of integration in eq.(\ref{I1}) is extended up to $\infty$ whereas for $\mathcal{I}^{\tilde{\nu}}_{2}$ in eq.(\ref{I2}), the upper limit of integration can not be extended to $\infty$ as the extra linear contribution in $y$ makes the value of the integrand to rise for higher values of $y$. One can now recast eq.(\ref{I1}) and execute the integration as
\begin{equation}\label{I1_Solved}
\begin{split}
\mathcal{I}^{\tilde{\nu}}_1\bigr\rvert_{y_f\rightarrow\infty}&=\int_0^\infty dy \exp[-iy]~ y^{-2i\tilde{\nu}}\\&=-2\tilde{\nu}e^{-\pi\tilde{\nu}}\Gamma[-2i\tilde{\nu}]~.
\end{split}
\end{equation}
The finite integral in eq.(\ref{I2}) reads
\begin{equation}\label{I2_Solved}
\begin{split}
\mathcal{I}_2^{\tilde{\nu}}&=\int_0^{y_f}dy~\exp[-iy]y^{1-2i\tilde{\nu}}\\
&=-e^{-\pi\tilde{\nu}}\gamma\left[2-2i\tilde{\nu},iy_f\right]\\
&=-e^{-\pi\nu \left(1+\frac{3\epsilon}{(\ln\xi)^2}\right)}\gamma\left[2-2i\nu \left(1+\frac{3\epsilon}{(\ln\xi)^2}\right),iy_f\right]
\end{split}
\end{equation}
where $\gamma\left[2-2i\tilde{\nu},iy_f\right]$ denotes the lower incomplete gamma function. One can represent the lower incomplete gamma function as
\begin{equation}\label{Gamma}
\gamma[2-2i\tilde{\nu},iy_f]=\Gamma[2
-2i\tilde{\nu}]-\Gamma[2-2i\tilde{\nu},iy_f]
\end{equation}
with $\Gamma[2-2i\tilde{\nu},iy_f]
$ denoting the upper-incomplete gamma function.
\begin{widetext}
\noindent As $y$ is small in the vicinity of the event horizon, one can, in reality, drop the contribution coming from the integral $\mathcal{I}_2^{\tilde{\nu}}$ from eq.(\ref{Probability_redefinition}) which leads to the Planckian spectrum given as
\begin{equation}\label{Planckianprob}
\begin{split}
\mathcal{P}_{\text{exc}}^{\text{Planck}} \simeq & \frac{4\pi \nu \mathcal{G}^{2}}{\Omega^{2}} \left[1+\frac{2\epsilon}{\ln{\xi}}+\frac{3\epsilon}{2(\ln{\xi})^2}-\frac{5\nu}{\Omega}\left(1+\frac{2\epsilon}{\ln{\xi}}\left(1+\frac{9}{10\ln{\xi}}-\frac{18}{5(\ln{\xi})^{2}}\right)\right)\right] \frac{1}{e^{4\pi\nu\left(1+\frac{3\epsilon}{(\ln{\xi})^{2}}\right)}-1}~.
\end{split}
\end{equation}
As can be seen from the excitation probability, it is Planckian in nature and as it is evident from the Planck factor, it has a direct contribution due to charge renormalization. 
From eq.(\ref{Planckianprob}), it is also easy to check that for very high field frequency $\nu$, the excitation probability will be very small. So we shall stick with the standard convention of taking $\nu\ll\Omega$ \cite{Scullypnas}. We can now write down the above expression with a proper dimensional reconstruction as
\begin{equation}\label{PlanckianDimension}
\begin{split}
\mathcal{P}_{\text{exc}}^{\text{Planck}}=&\frac{4\pi\nu\mathcal{G}^2}{\Omega^2}\left[\frac{2 G_0 M}{c^3}+\frac{\hbar e_*^2}{M\tilde{\alpha} c^2\ln\left[\frac{8\pi \tilde{\alpha}G_0 M^2}{\sqrt{3}\chi^2\hbar c}\right]}+\frac{3\hbar e_*^2}{4M\tilde{\alpha} c^2\left(\ln\left[\frac{8\pi \tilde{\alpha}G_0 M^2}{\sqrt{3}\chi^2\hbar c}\right]\right)^2}-\frac{5\nu}{\Omega}\left[\frac{2 G_0 M}{c^3}+\frac{\hbar e_*^2}{M\tilde{\alpha} c^2\ln\left[\frac{8\pi \tilde{\alpha}G_0 M^2}{\sqrt{3}\chi^2\hbar c}\right]}\right.\right.\\\times&\left.\left.\left(1+\frac{9}{10\ln\left[\frac{8\pi G_0 M^2}{\sqrt{3}\chi^2\hbar c}\right]}-\frac{18}{5\left(\ln\left[\frac{8\pi G_0 M^2}{\sqrt{3}\chi^2\hbar c}\right]\right)^2}\right)\right]\right]\frac{1}{\exp\left[4\pi\nu\left(\frac{2G_0M}{c^3}+\frac{3\hbar e_*^2}{2M\tilde{\alpha} c^2\left(\ln\left[\frac{8\pi \tilde{\alpha}G_0 M^2}{\sqrt{3}\chi^2\hbar c}\right]\right)^2}\right)\right]-1}
\end{split}
\end{equation}
where the restoration is done considering the fact that $e_*^2$ denotes the dimensionless charge parameter which can be expressed in terms of the dimensionful charge cut-off $e_*^d$ as $e_*^2\equiv \frac{{e_*^d}^2G_0}{4\pi \varepsilon_0 c^6}$ with $\varepsilon_0$ denoting the permittivity of free space.
\end{widetext}

\noindent We shall now explore the complete structure of the excitation probability by taking into effect the analytical form of the integral in eq.(\ref{I2_Solved}). The analytical form of the excitation probability in eq.(\ref{Probability_redefinition}), can be recast in the form given as
\begin{equation}\label{excprob}
    \begin{split}
        \mathcal{P}_{\text{exc}} = &\frac{\mathcal{G}^{2}\mathcal{A}_\nu}{\Omega^{2}}  \biggr|-2\pi \tilde{\nu} \left[1+\frac{\epsilon}{\log{\xi}}\right]e^{-\pi \tilde{\nu}}\Gamma[-2i\tilde{\nu}] \\ -& \frac{\mathcal{B}_\nu}{2\Omega}e^{-\pi \bar{\nu}}\gamma[2-2i\tilde{\nu},iy_{f}]\biggr|^{2}~.
    \end{split}
\end{equation}

\noindent We shall now make use of the following relations to simplify the form of the excitation probability above
\begin{equation}\label{identity2}
\begin{split}
    \Gamma[-2i\tilde{\nu}]&=\left(\frac{\pi}{2\tilde{\nu}\sinh{(2\pi\tilde{\nu})}}\right)^{\frac{1}{2}}e^{i \text{arg}[\Gamma[-2i\tilde{\nu}]]}\\
    |\Gamma[-2i\tilde{\nu}]|^{2} &= \frac{\pi}{2\tilde{\nu}\sinh{(2\pi\tilde{\nu})}}\\ \text{arg}[\Gamma[-2i\tilde{\nu}]]&=-\text{arg}[\Gamma[2i\tilde{\nu}]]
    \end{split}
\end{equation}
where the last identity comes form the fact that $\Gamma[2i\tilde{\nu}]$ is complex conjugate to $\Gamma[2i\tilde{\nu}]$. Using the relations in eq.(\ref{identity2}) along with eq.(\ref{Planckianprob}), one can write down the form of the transition probability from eq.(\ref{excprob}) as
\begin{widetext}
\begin{equation}\label{totalexcprob}
\begin{split}
        \mathcal{P}_{\text{exc}} = & \mathcal{P}_{\text{exc}}^{\text{Planck}} + \frac{\mathcal{G}^{2}\pi^{2}\sqrt{\tilde{\nu}}\left(1+\frac{\epsilon}{\log{\xi}}\right)}{\Omega^{3}}\mathcal{A}_\nu\mathcal{B}_\nu e^{-\pi\tilde{\nu}}\left[\frac{e^{-i\text{arg}[\Gamma[2i\tilde{\nu}]]}}{\sqrt{e^{4\pi\tilde{\nu}}-1}}\gamma(2+2i\tilde{\nu},-iy_{f}) +\frac{e^{i\text{arg}[\Gamma[2i\tilde{\nu}]]}}{\sqrt{e^{4\pi\tilde{\nu}}-1}}\gamma[2-2i\tilde{\nu},iy_{f}]\right] \\&+\frac{\mathcal{G}^{2}}{4\Omega^{4}}\mathcal{A}_\nu\mathcal{B}_\nu^{2}e^{-2\pi\tilde{\nu}}\gamma[2+2i\tilde{\nu},-iy_{f}] \gamma[2-2i\tilde{\nu},iy_{f}]~.
\end{split}
\end{equation}
\end{widetext}
This total excitation probability (\ref{totalexcprob}) consists of some deformed Planckian behaviour due to the presence of incomplete gamma functions. These terms will contribute to the overall probability only for nonvanishing values of $y_{f}$ \cite{kerr2} and hence the emitted radiation from the atom will not be black-body type anymore, in place, it will be slightly deformed in nature. 
\begin{widetext}
\par Now, one can obtain the overall absorption probability by substituting $-\nu$ in place of $\nu$ as following
\begin{equation}\label{totalexcprob}
\begin{split}
        \mathcal{P}_{\text{abs}} = & \mathcal{P}_{\text{abs}}^{\text{Planck}} + \frac{\mathcal{G}^{2}\pi^{2}\sqrt{\tilde{\nu}}\left(1+\frac{\epsilon}{\log{\xi}}\right)}{\Omega^{3}}\mathcal{A}_{-\nu}\mathcal{B}_{-\nu} e^{\pi\tilde{\nu}}\left[\frac{e^{i\text{arg}[\Gamma[2i\tilde{\nu}]]}}{\sqrt{1-e^{-4\pi\tilde{\nu}}}}\gamma(2-2i\tilde{\nu},-iy_{f}) +\frac{e^{-i\text{arg}[\Gamma[2i\tilde{\nu}]]}}{\sqrt{1-e^{-4\pi\tilde{\nu}}}}\gamma[2+2i\tilde{\nu},iy_{f}]\right] \\&+\frac{\mathcal{G}^{2}}{4\Omega^{4}}\mathcal{A}_{-\nu}\mathcal{B}_{-\nu}^{2}e^{2\pi\tilde{\nu}}\gamma[2-2i\tilde{\nu},-iy_{f}] \gamma[2+2i\tilde{\nu},iy_{f}]
\end{split}
\end{equation}
where $\mathcal{P}_{\text{abs}}^{\text{Planck}}$ can be obtained from eq.(\ref{Planckianprob}) (by using $-\nu$ instead of $\nu$) as
\begin{equation}\label{Planckianabsprob}
 \begin{split}
 \mathcal{P}_{\text{abs}}^{\text{Planck}} \simeq & \frac{4\pi \nu \mathcal{G}^{2}}{\Omega^{2}} \left[1+\frac{2\epsilon}{\ln{\xi}}+\frac{3\epsilon}{2(\ln{\xi})^2}+\frac{5\nu}{\Omega}\left(1+\frac{2\epsilon}{\ln{\xi}}\left(1+\frac{9}{10\ln{\xi}}-\frac{18}{5(\ln{\xi})^{2}}\right)\right)\right] \frac{1}{1-e^{-4\pi\nu\left(1+\frac{3\epsilon}{(\ln{\xi})^{2}}\right)}}~.
\end{split}
\end{equation} 
One can once again write down the absorption probability with dimensional reconstruction as
\begin{equation}\label{PlanckianAbsorptionDimension}
\begin{split}
\mathcal{P}_{\text{abs}}^{\text{Planck}}=&\frac{4\pi\nu\mathcal{G}^2}{\Omega^2}\left[\frac{2 G_0 M}{c^3}+\frac{\hbar e_*^2}{M\tilde{\alpha} c^2\ln\left[\frac{8\pi \tilde{\alpha}G_0 M^2}{\sqrt{3}\chi^2\hbar c}\right]}+\frac{3\hbar e_*^2}{4M\tilde{\alpha} c^2\left(\ln\left[\frac{8\pi \tilde{\alpha}G_0 M^2}{\sqrt{3}\chi^2\hbar c}\right]\right)^2}+\frac{5\nu}{\Omega}\left[\frac{2 G_0 M}{c^3}+\frac{\hbar e_*^2}{M\tilde{\alpha} c^2\ln\left[\frac{8\pi \tilde{\alpha}G_0 M^2}{\sqrt{3}\chi^2\hbar c}\right]}\right.\right.\\\times&\left.\left.\left(1+\frac{9}{10\ln\left[\frac{8\pi G_0 M^2}{\sqrt{3}\chi^2\hbar c}\right]}-\frac{18}{5\left(\ln\left[\frac{8\pi G_0 M^2}{\sqrt{3}\chi^2\hbar c}\right]\right)^2}\right)\right]\right]\frac{1}{1-\exp\left[-4\pi\nu\left(\frac{2G_0M}{c^3}+\frac{3\hbar e_*^2}{2M\tilde{\alpha} c^2\left(\ln\left[\frac{8\pi \tilde{\alpha}G_0 M^2}{\sqrt{3}\chi^2\hbar c}\right]\right)^2}\right)\right]}~.
\end{split}
\end{equation}
\end{widetext}
\noindent To calculate the horizon brightened acceleration radiation entropy, we shall be using the Planckian part of the probabilities and will ignore the non-Planckian part for better comparison with earlier analyses.
It is important to understand that the contributions due to the lower incomplete gamma functions are very small in the excitation probability eq.(\ref{totalexcprob}) and as a result, the deviation from the Planckian behaviour is rather insignificant. As a result, one can disregard the contributions due to the non-Planckian part in the HBAR entropy which will be calculated in the next section.
\section{Calculation of the HBAR entropy}\label{S4}
\noindent The concept of horizon brightened acceleration radiation (HBAR) entropy was first introduced in \cite{Scullypnas} to distinguish between the Bekenstein-Hawking entropy and the entropy due to the infalling atoms into the event horizon of the black hole. In this section, we are going to calculate how the HBAR entropy is being modified with the flow of couplings in the quantum-improved charged black hole spacetime. We assume the same thought experiment of a two-level atom with angular frequency $\Omega$ which is freely falling into the event horizon of a quantum-improved charged black hole with a constant rate of fall $\kappa_f$. To find the von Neumann entropy we shall make use of the density matrix formalism. If the microscopic change in the field density matrix due to one atom is $\delta \rho_{i}$, then for $\Delta N$ number of atoms, the total microscopic change of the photon density matrix will be
\begin{equation}
    \Delta \rho = \sum_{i} \delta \rho_{i} = \delta \rho \Delta N
\end{equation}
where $\Delta N = \kappa_f \Delta t$ and the underlying assumption is that $\delta\rho_i=\delta\rho$ $\forall i\in [1,\Delta N]$. The equation of motion then takes the simple form 
\begin{equation}
    \frac{\Delta \rho}{\Delta t} = \kappa_f \delta \rho~.
\end{equation}
Now, for an arbitrary field state $|n\rangle$ which denotes a state with $n$ scalar photons, the time rate of change of the $\{n,n\}$ component of the radiation density matrix is given by 
\begin{align}\label{eom}
    \dot{\rho}_{n,n} = & -\Gamma_{\text{abs}} (n\rho_{n,n}-(n+1)\rho_{n+1,n+1})\nonumber \\&-\Gamma_{\text{exc}} ((n+1)\rho_{n,n}-n\rho_{n-1,n-1})
\end{align}
where $\Gamma_{\text{exc}}$ and $\Gamma_{\text{abs}}$ are the excitation and absorption rate respectively and they are given by $\Gamma_{\text{exc}/\text{abs}}=\kappa_f \mathcal{P}_{\text{exc}/\text{abs}}$. The steady-state solution corresponding to eq.(\ref{eom}) will be used to calculate the HBAR entropy. For a steady state scenario $\dot{\rho}_{n,n}$ will vanish and one can get the relation between the $\{0,0\}$ and $\{1,1\}$ elements of the density matrix as
    $\rho_{1,1}=\frac{\Gamma_{exc}}{\Gamma_{abs}}\rho_{0,0}$. Doing the iteration for $n$ times, one can get the recursion relation as 
\begin{equation}\label{recursion_density}
    \rho_{n,n}=\left(\frac{\Gamma_{\text{exc}}}{\Gamma_{\text{abs}}}\right)^{n}\rho_{0,0}~.
\end{equation}
Now, we need to find the $\rho_{0,0}$ component of the density matrix to obtain the steady-state solution. Using the property of density matrix $Tr(\rho) = 1$, one obtains the analytical form of the $\{0,0\}$ component of the density matrix as $\rho_{0,0}= 1- \frac{\Gamma_{\text{exc}}}{\Gamma_{\text{abs}}}$. Hence, the steady state solution can finally be obtained as 
\begin{equation}\label{steadystate}
    \rho_{n,n}^{s.s.}=\left(\frac{\Gamma_{\text{exc}}}{\Gamma_{\text{abs}}}\right)^{n}\left(1- \frac{\Gamma_{\text{exc}}}{\Gamma_{\text{abs}}}\right)~.
\end{equation}
We shall now make use of the Planckian part of the excitation probability from eq.(\ref{Planckianprob}) and also for the absorption probability in eq.(\ref{Planckianabsprob}) to get the ratio of the excitation and the absorption rate as
\begin{equation}\label{ratio}
\begin{split}
     \frac{\Gamma_{\text{exc}}}{\Gamma_{\text{abs}}} \simeq &\left[ 1-\frac{10\nu}{\Omega}+\frac{12\epsilon\nu}{\Omega(\ln{\xi})^{2}}\left[1+\frac{6}{\ln{\xi}}\right]\right]  e^{-4\pi\nu\left[1+\frac{3\epsilon}{(\ln{\xi})^{2}}\right]}.
\end{split}
\end{equation}
With the analytical form of $\frac{\Gamma_{\text{exc}}}{\Gamma_{\text{abs}}}$ obtained in the above equation one can obtain the $\{n,n\}$ component of the steady state solution of the density matrix from eq.(\ref{steadystate}). The rate of change of the von Neumann entropy takes the form given as
\begin{equation}\label{vonNeumann}
    \dot{S_{\rho}} = -K_{B}\sum_{n,\nu} \dot{\rho}_{n,n}\ln{(\rho_{n,n})}~.
\end{equation}
Following \cite{Scullypnas}, we can replace $\rho_{n,n}$ inside of the logarithmic term and replace it with the $\{n,n\}$ component of the steady state density matrix. Hence, we can recast eq.(\ref{vonNeumann}) as
\begin{equation}\label{entropychange}
    \dot{S_{\rho}} = -K_{B}\sum_{n,\nu} \dot{\rho}_{n,n}\ln{(\rho_{n,n}^{s.s.})}~.
\end{equation}
Using the analytical form of the steady state solution of the density matrix and using the relation $\sum_n n\dot{\rho}_{n,n}=\frac{d}{dt}\left(\sum_n n\rho_{n,n}\right)=\dot{\bar{n}}_\nu$, we arrive at the relation for the rate of change of the von-Neumann entropy as
\begin{widetext}
\begin{equation}\label{ent1}
\begin{split}
    \dot{S_{\rho}} \simeq &4\pi K_{B}\left(1+\frac{3\epsilon}{(\ln{\xi})^{2}}\right)\sum_{\nu}\dot{\bar{n}}_{\nu}\nu+ \frac{K_{B}}{\Omega}\left(10-\frac{12\epsilon}{(\ln{\xi})^{2}}\left(1+\frac{6}{\ln{\xi}}\right)\right) \sum_{\nu}\dot{\bar{n}}_{\nu}\nu
\end{split}
\end{equation}
where $\dot{\bar{n}}_{\nu}$ is the flux due to the emission of photons from infalling atoms and the energy loss due to this emission of photons reads $\hbar\sum_\nu \dot{\bar{n}}_{\nu} \nu = \dot{m}_{p}c^{2}$. Using this relation and restoring dimensions, we can recast eq.(\ref{ent1}) as
\begin{align}\label{ent2}
    \dot{S_{\rho}} \simeq &\frac{4\pi K_{B} \dot{m}_{p}c^{2} }{\hbar}\left[\frac{2G_{0}M}{c^{3}}+\frac{3e_{*}^{2}\hbar}{2\tilde{\alpha}c^{2}M\left[\ln{\left[\frac{8\pi\tilde{\alpha}G_{0}M^{2}}{\sqrt{3}\chi^{2}\hbar c}\right]}\right]^{2}}\right]+ \frac{K_{B}\dot{m}_{p}c^{2}}{\hbar\Omega}\left[ 10-\frac{3e_{*}^{2}c\hbar}{\tilde{\alpha}G_{0}M^{2}\left[\ln{\left[\frac{8\pi\tilde{\alpha}G_{0}M^{2}}{\sqrt{3}\chi^{2}\hbar c}\right]}\right]^{2}}\left[1+\frac{6}{\ln{[\frac{8\pi\tilde{\alpha}G_{0}M^{2}}{\sqrt{3}\chi^{2}\hbar c}]}}\right]\right].
\end{align}
\end{widetext}
Now, we aim to find the rate of change of entropy in terms of the area of the quantum-improved charged black hole. Using proper dimensional reconstruction, we can rewrite the analytical form of the event horizon of the black hole from eq.(\ref{horizon}) as
\begin{equation}
    r_{+} = \frac{2G_{0}M}{c^{2}}- \frac{\hbar e_{*}^{2}}{2\tilde{\alpha}Mc\ln{[\frac{8\pi\tilde{\alpha}G_{0}M^{2}}{\sqrt{3}\chi^{2}\hbar c}]}}~.
\end{equation}
In terms of the event horizon for the black hole, we can rewrite the area of the quantum-improved charged black hole black hole (QICBH) as
\begin{equation}\label{area}
    A_{\text{QICBH}} = 4\pi r_{+}^{2}\simeq \frac{16\pi G_{0}^{2}M^{2}}{c^{4}}- \frac{8\pi \hbar e_{*}^{2}G_{0}}{c^{3}\tilde{\alpha}\ln{\left[\frac{8\pi\tilde{\alpha}G_{0}M^{2}}{\sqrt{3}\chi^{2}\hbar c}\right]}}~.
\end{equation}
As via infall of atoms and emission of photons, the mass of the black hole changes with time, the time derivative of the mass of the black hole will not vanish. Taking a time derivative of both sides of the above equation, we obtain the rate of change of the area of the black hole as
\begin{equation}\label{area_change}
    \dot{A}_{\text{QICBH}} \simeq \frac{32\pi G_{0}^{2}M\dot{M}}{c^{4}} + \frac{16\pi \hbar e_{*}^{2}G_{0}\dot{M}}{c^{3}M\tilde{\alpha}\left(\ln{\left[\frac{8\pi\tilde{\alpha}G_{0}M^{2}}{\sqrt{3}\chi^{2}\hbar c}\right]}\right)^{2}}
\end{equation}
where the total rate of change of the mass of the black hole, $\dot{M}$, is defined as the sum of the rate of change of mass due to the emission of photons and addition of infalling atoms in the event horizon which is expressed analytically as $\dot{M}=\dot{m}_{p}+\dot{m}_{\text{atom}}$. So, when there is no atom falling into the event horizon, then the change of the area due to photon emission $\dot{A}_{p}$ will be the same as the change of the area of the black hole. Hence, we can express the rate of change of the area of the black hole due to the emission of photons as
\begin{align}\label{ap}
    \dot{A}_{p} \simeq \frac{32\pi G_{0}^{2}M\dot{m}_{p}}{c^{4}} + \frac{16\pi \hbar e_{*}^{2}G_{0}\dot{m}_{p}}{c^{3}M\tilde{\alpha}\left(\ln{\left[\frac{8\pi\tilde{\alpha}G_{0}M^{2}}{\sqrt{3}\chi^{2}\hbar c}\right]}\right)^{2}}~.
\end{align}
Using the above expression, we can rewrite eq.(\ref{ent2}) as
\begin{widetext}
\begin{align}\label{ent3}
    \dot{S}_{\rho} \simeq &\frac{K_{B}c^{3}\dot{A_{p}}}{4\hbar G_{0}}+\frac{2\pi K_{B}e_{*}^{2}c^{3}}{\tilde{\alpha}\left(\ln{\left[\frac{8\pi\tilde{\alpha}G_{0}M^{2}}{\sqrt{3}\chi^{2}\hbar c}\right]}\right)^{2}}\left(\frac{\dot{m}_{p}}{M}\right)+\frac{K_{B}}{\hbar\Omega}\left[ 10-\frac{3e_{*}^{2}c\hbar}{\tilde{\alpha}G_{0}M^{2}\left(\ln{\left[\frac{8\pi\tilde{\alpha}G_{0}M^{2}}{\sqrt{3}\chi^{2}\hbar c}\right]}\right)^{2}}\left[1+\frac{6}{\ln{[\frac{8\pi\tilde{\alpha}G_{0}M^{2}}{\sqrt{3}\chi^{2}\hbar c}]}}\right]\right] \dot{m}_{p}c^{2}~.
\end{align}
\end{widetext}
\pagebreak

\begin{figure}[ht!]
\begin{center}
\includegraphics[scale=0.345]{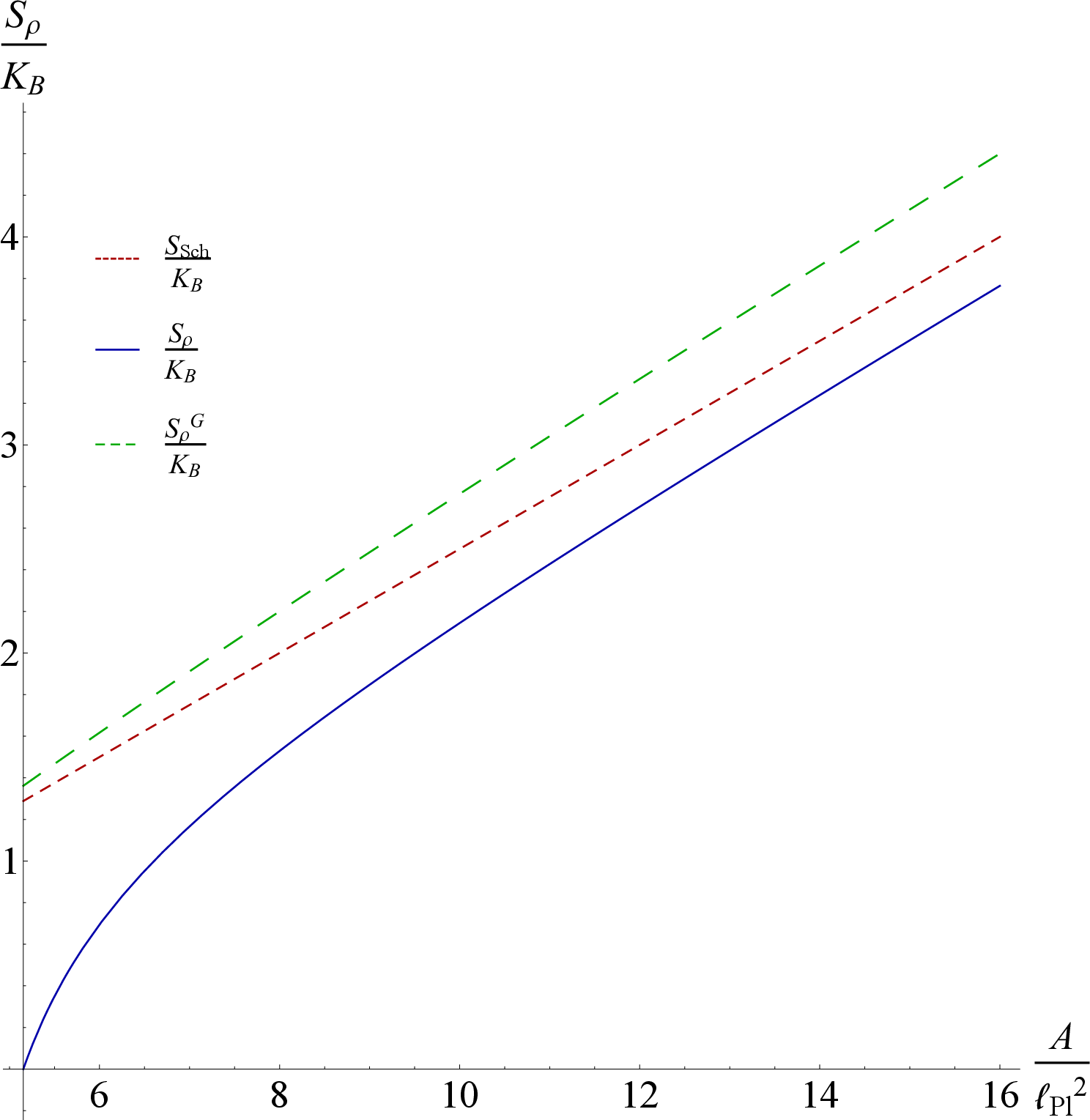}
\caption{Dimensionless entropy versus dimensionless area for the general Schwarzschild black hole, quantum-improved charged black hole with both $G$ and $e$ flow, and quantum-improved charged black hole with only the flow of the gravitational constant\label{Entropy_Plot}.}
\end{center}
\end{figure}

Earlier, we expressed the event horizon $r_{+}$ of the black hole in terms of the Schwarzschild radius and the small quantum corrected term. Hence, we can also express the area of the black hole in a similar way, $A_{\text{QICBH}} = A_{\text{Sch}} + \epsilon A_{\text{QC}}$, where $A_{\text{Sch}}=4\pi r_{\text{Sch}}^{2}=\frac{16\pi G_{0}^{2}M^{2}}{c^{4}}$. Hence, from eq.(\ref{area}), one can express the mass of the black hole in terms of its event horizon area $A_{\text{QICBH}}$ as
\begin{equation}\label{Meq}
\begin{split}
    A_{\text{QICBH}} = &A_{\text{Sch}} - \frac{8\pi \hbar e_{*}^{2}G_{0}}{c^{3}\tilde{\alpha}\ln{\left[\frac{\tilde{\alpha}c^{3}A_{\text{Sch}}}{2\sqrt{3}\chi^{2}\hbar G_{0}}\right]}} \\
    \Rightarrow A_{\text{Sch}}\simeq & A_{\text{QICBH}} + \frac{8\pi \hbar e_{*}^{2}G_{0}}{c^{3}\tilde{\alpha}\ln{\left[\frac{\tilde{\alpha}c^{3}A_{\text{QICBH}}}{2\sqrt{3}\chi^{2}\hbar G_{0}}\right]}} ~.  
\end{split}
\end{equation}
Substituting the form of $A_{\text{Sch}}$ in the above equation, we can write down the relation between the mass of the black hole and $A_p$ as
\begin{equation}\label{Mass_Relation}
\frac{16\pi G_{0}^{2}M^{2}}{c^{4}} =  A_{p}+\frac{8\pi \hbar e_{*}^{2}G_{0}}{c^{3}\tilde{\alpha}\ln{\left[\frac{\tilde{\alpha}c^{3}A_{p}}{2\sqrt{3}\chi^{2}\hbar G_{0}}\right]}}~.
\end{equation}
\noindent Differentiating both sides of eq.(\ref{Meq}) with respect to time and substituting the analytical form of $\dot{m}_p$ into eq.(\ref{ent3}), the rate of change of entropy of the black hole takes the form in terms of the rate of change of the area of the black hole as
\begin{widetext}
\begin{equation}\label{Entropy}
\begin{split}
    \dot{S}_{\rho} \simeq &\frac{K_{B}c^{3}\dot{A_{p}}}{4\hbar G_{0}} - \frac{\pi K_{B}e_{*}^{2}}{\tilde{\alpha}}\frac{d}{dt}\left(\frac{1}{\ln{\left[\frac{\tilde{\alpha}c^{3}A_{p}}{2\sqrt{3}\chi^{2}\hbar G_{0}}\right]}}\right) + \frac{5K_{B}c^{4}}{2\sqrt{\pi}\hbar \Omega G_{0}}\dot{A}_{p}^{1/2}\nonumber \\
    = & \frac{d}{dt} \left[\frac{K_{B}c^{3}A_{p}}{4\hbar G_{0}} - \frac{\pi K_{B}e_{*}^{2}}{\tilde{\alpha}\log{[\frac{\tilde{\alpha}c^{3}A_{p}}{2\sqrt{3}\chi^{2}\hbar G_{0}}]}} + \frac{5K_{B}c^{4}}{2\sqrt{\pi}\hbar \Omega G_{0}}A_{p}^{1/2}\right]~.
    \end{split}
\end{equation}
\end{widetext}
This equation tells about how the entropy is related to the area of a quantum-improved charged black hole geometry. In this case, the leading term of this HBAR entropy is as usual following the ``\textit{Area divided by four}" law\cite{Scullypnas}. The most important observation in our current analysis lies in the fact that unlike \cite{JanaSenGangopadhyay}, we observe an inverse logarithmic correction in the area of the black hole. Along with the standard ``area divided by four" law and the inverse logarithmic correction, we also observe a square root of area correction in the entropy term. This inverse logarithmic correction to the HBAR entropy is a completely new result and to top it all, we observe that it comes with an overall minus sign. This behaviour implies that this inverse logarithmic correction reduces the overall standard entropy of the black hole and with the increase in the mass of the black hole this inverse logarithmic correction becomes smaller. Hence, for the most effective contribution from this kind of term, one needs to investigate black holes with smaller masses. 
This is the most important finding in our work.
We shall now plot this inverse entropy correction case with the case obtained earlier in \cite{JanaSenGangopadhyay}. The form of the entropy obtained earlier for a black hole with flowing Newtonian gravitational constant only had the form
\begin{equation}\label{QCBH}
S_\rho^{G}=\frac{K_Bc^3A_p}{4\hbar G_0}+\pi \tilde{\omega} K_B\log\left[\frac{A_p c^3}{4\hbar G_0}\right]
\end{equation}
where the suffix $G$ in $S_\rho$ denotes that only the gravitational constant flows with the momentum scale and $\tilde{\omega}$ is the quantum gravity parameter. Ignoring the $A_p^{\frac{1}{2}}$ term in eq.(\ref{Entropy}), we get the form of the entropy for our current quantum-improved charged black hole as
\begin{equation}\label{QICBH_Entropy}
S_\rho=\frac{K_{B}c^{3}A_{p}}{4\hbar G_{0}} - \frac{\pi K_{B}e_{*}^{2}}{\tilde{\alpha}\log{[\frac{\tilde{\alpha}c^{3}A_{p}}{2\sqrt{3}\chi^{2}\hbar G_{0}}]}}~.
\end{equation}
If $\chi=1$, then for simpler structure of the entropy, we can set $\tilde{\alpha}=\frac{\sqrt{3}}{2}$ which is greater than zero and less than unity. For this value of $\tilde{\alpha}$, $\tilde{\omega}=\frac{1}{4\pi\tilde{\alpha}}=\frac{1}{2\sqrt{3}\pi}<1$. It is important to observe that we cannot consider black holes with sufficiently small radii as for eq.(\ref{QICBH_Entropy}), the entropy can be negative for a certain value of the dimensionless quantity $\frac{A}{l_{Pl}^2}$. For the value of the dimensionless charge parameter $e_*=0.3$, we get the lower bound to the area of the black hole to be of the order of $A\simeq5.15~ l_{Pl}^2.$ 
Some important comments are in order. The phenomena of the entropy of the black hole becoming negative is very interesting. The physical implication is that a charged static black hole where the charge as well as Newton's gravitational constant flows with the radial distance can not sustain a steady state (for this current system) if the area of the black hole is below a certain lower bound.
We shall now plot the entropy divided by the Boltzmann constant against $\frac{A}{l_{Pl}^2}$ for eq.(s)(\ref{QCBH},\ref{QICBH_Entropy}) along with the HBAR entropy for standard Schwarzschild black hole in Fig.(\ref{Entropy_Plot}).
As we can see from Fig.(\ref{Entropy_Plot}), the entropy for the current analysis is even smaller than the HBAR entropy for a standard Schwarzschild black hole \cite{Scullypnas}. This observation implies that the renormalization group improved black hole harbours a lower amount of disorder than the standard Schwarzschild as well as the quantum-improved charged black hole with only the flow of Newton's gravitational constant. This is a very important observation in our analysis.
\section{conclusion}\label{S5}
\noindent In this paper, we have considered the standard thought experiment of a stream of two-level atoms freely falling into the event horizon of a quantum corrected non-rotating charged black hole where the event horizon of the black hole is shielded by a mirror to protect the infalling atoms from interacting with the Hawking radiation. This mirror also ensures that the vacuum states of the scalar field are Boulware vacuum corresponding to an asymptotic observer. Following the discussion in \cite{Ishibashi}, we have considered the case of running couplings for the charge of the black hole as well as Newton's gravitational constant. We consider the structure of the lapse function in the $r\gg 0$ case as we have worked in the near horizon of the black hole. As the atoms freely fall into the event horizon of the black hole it gets excited and simultaneously emits a virtual photon. We have calculated the excitation probability corresponding to the atom going from its ground state to the first excited energy state with the simultaneous emission of a virtual photon and observed that the leading order Planck factor has a contribution from the charge renormalization. We also consider a spherically symmetric solution of the scalar field and observed that the higher order contribution deforms the Planckian spectrum. The deformation of Planckian nature is due to the finite value of the upper limit of the integration in $\mathcal{I}_{2}^{\tilde{\nu}}$, which gives rise to the incomplete gamma functions. Before falling into the black hole, the finite path covered by the atoms is responsible for this deformed nature. Finally, we have calculated the HBAR entropy for this acceleration-radiation phenomenon using the density matrix formalism in quantum statistical mechanics. We observe that the leading order term in the von Neumann entropy follows the standard ``area divided by four" law whereas there are subleading inverse logarithmic corrections in the area of the black hole and a square root correction in the area of the black hole. The inverse logarithmic correction is a completely new type of correction and has a complete quantum gravitational origin. We observe that the inverse logarithmic term comes with an overall minus sign which states that for black holes with smaller masses, the entropy becomes very small and for micro black holes such a term will have a significant contribution.
Finally, we have plotted the dimensionless HBAR entropy of the black hole $\frac{S_\rho}{K_B}$ against the dimensionless area of the black hole given by $\frac{A}{l_{Pl}^2}$. We observe that the HBAR entropy is lower than the standard Schwarzschild black hole case and asymptotically approaches the ``area divided by four" value for higher values of the event horizon area of the black hole. This implies that the measure of disorder is lower for this current renormalization group improved charged black hole with flowing Newton's gravitational constant and black hole charge parameter.


\begin{thebibliography}{8}
\bibitem{EinsteinGR1}
A. Einstein, ``\textit{Die Grundlage der allgemeinen Relativit\"{a}tstheorie}", \href{https://onlinelibrary.wiley.com/doi/10.1002/andp.19163540702}{Ann. der Phys. (Berlin) 354 (1916) 769}.
\bibitem{EinsteinGR2}
A. Einstein, ``\textit{Die Feldgleichungen der Gravitation}", Sitz. Preu\ss. Akad. Wiss. 844 (1915).
\bibitem{Reuter1}
M. Reuter, "\textit{Nonperturbative evolution equation for quantum gravity}", \href{ https://doi.org/10.1103/PhysRevD.57.971}{Phys. Rev. D 57 (1998) 971}.
\bibitem{Reuter2}
M. Reuter and F. Saueressig, "\textit{Quantum Gravity and the Functional Renormalization Group : The Road towards Asymptotic Safety}", (Cambridge University Press, Cambridge, England, 2019).
\bibitem{Percacci}
R. Percacci, "\textit{An Introduction to Covariant Quantum Gravity and Asymptotic Safety}",(World Science, Singapore,2017).
\bibitem{BonannoReuter}
A. Bonanno and M. Reuter, ``\textit{Renormalization group improved black hole spacetimes}", \href{https://doi.org/10.1103/PhysRevD.62.043008}{Phys. Rev. D 62 (2000) 043008}.
\bibitem{BonannoReuter2}
A. Bonanno and M. Reuter, ``\textit{Spacetime structure of an evaporating black hole in quantum gravity}", \href{ https://doi.org/10.1103/PhysRevD.73.083005}{Phys. Rev. D 73 (2006) 083005}.
\bibitem{ReuterTuiran}
M. Reuter and E. Tuiran, ``\textit{Quantum gravity effects in the Kerr spacetime}", \href{https://doi.org/10.1103/PhysRevD.83.044041}{Phys. Rev. D 83 (2011) 044041}.
\bibitem{Falls}
K. Falls, D. F. Litim and A. Raghuraman, ``\textit{Black holes and asymptotically safe gravity}", \href{https://doi.org/10.1142/S0217751X12300037}{Int. J. Mod. Phys. A 27 (2012) 1250019}.
\bibitem{Koch1}
B. Koch and F. Saueressig, ``\textit{Structural aspects of asymptotically safe black holes}", \href{https://doi.org/10.1088/0264-9381/31/1/015006}{Class. Quantum Grav. 31 (2014) 015006}.
\bibitem{Koch2}
B. Koch and F. Saueressig, ``\textit{Black holes within asymptotic safety}", \href{https://doi.org/10.1142/S0217751X14300117}{ Int. J. Mod. Phys. A 29 (2014) 1430011}.
\bibitem{BonannoKoch}
 A. Bonanno, B. Koch, and A. Platania, ``\textit{Gravitational Collapse in Quantum Einstein Gravity}", \href{https://doi.org/10.1007/s10701-018-0195-7}{Found. Phys. 48 (2018) 1393}. 
\bibitem{PawlowskiStock}
J. M. Pawlowski and D. Stock, ``\textit{Quantum-improved Schwarzschild-(A)dS and Kerr-(A)dS spacetimes}", \href{https://doi.org/10.1103/PhysRevD.98.106008}{Phys. Rev. D 98 (2018) 106008}.
\bibitem{Platania1}
A. Platania, ``\textit{Dynamical renormalization of black-hole spacetimes}", \href{https://doi.org/10.1140/epjc/s10052-019-6990-2}{Eur. Phys. J. C 79 (2019) 470}.
\bibitem{BonannoCasadio}
A. Bonanno, R. Casadio and A. Platania, ``\textit{Gravitational antiscreening in stellar interiors}", \href{https://doi.org/10.1088/1475-7516/2020/01/022}{J. Cosmol. Astropart. Phys. 01 (2020) 022}.
\bibitem{Platania2}
A. Platania, ``\textit{From Renormalization Group Flows to Cosmology}", \href{https://doi.org/10.3389/fphy.2020.00188 }{Front. Phys. 8 (2020) 188}.
\bibitem{ReuterWeyer}
M. Reuter and H. Weyer, ``\textit{Running Newton constant, improved gravitational actions, and galaxy rotation curves}", \href{ https://doi.org/10.1103/PhysRevD.70.124028}{Phys. Rev. D 70 (2004) 124028}.
\bibitem{Ishibashi}
A. Ishibashi, N. Ohta and D. Yamaguchi, ``\textit{Quantum improved charged black holes}", \href{https://doi.org/10.1103/PhysRevD.104.066016}{Phys. Rev. D 104 (2021) 06601}.
\bibitem{HarstReuter}
U. Harst and M. Reuter, ``\textit{QED coupled to QEG}",\href{https://doi.org/10.1007/JHEP05(2011)119}{J. High Energy Phys. 05 (2011) 119}.
\bibitem{Christiansen}
N. Christiansen and A. Eichhorn, ``\textit{An asymptotically safe solution to the U(1) triviality problem}", \href{https://doi.org/10.1016/j.physletb.2017.04.047}{Phys. Lett. B 770 (2017) 154}.
 \bibitem{Eichhorn}
A. Eichhorn and F. Versteegen, ``\textit{Upper bound on the Abelian gauge coupling from asymptotic safety}", \href{https://link.springer.com/article/10.1007/JHEP01(2018)030}{J. High Energy Phys. 01 (2018) 030}.
\bibitem{JanaSenGangopadhyay}
A. Jana, S. Sen, and S. Gangopadhyay, `` \textit{Atom falling into a quantum corrected  charged black hole and HBAR entropy}", \href{https://doi.org/10.1103/PhysRevD.110.026029}{Phys. Rev. D 110 (2024) 026029}.
\bibitem{RuizTuiran}
O. Ruiz and E. Tuiran, ``\textit{Nonperturbative quantum correction to the Reissner-Nordstr\"{o}m spacetime with running Newton’s constant}",\href{https://doi.org/10.1103/PhysRevD.107.066003}{Phys. Rev. D 107 (2023) 066003}.

\bibitem{Scullypnas}
M. O. Scully, S. Fulling, D. M. Lee, D. N. Page, W. P. Schleich, and A. A. Svidzinsky, ``\textit{Radiation from Atoms Falling into a Black Hole}", \href{https://doi.org/10.1073/pnas.1807703115}{Proc. Natl. Acad. Sci. 115 (2018) 8131}.
\bibitem{Hawking1}
S. W. Hawking, ``\textit{Black hole explosions?}", \href{https://doi.org/10.1038/248030a0}{Nature 248 (1974) 30}.
\bibitem{Hawking2}
S. W. Hawking, ``\textit{Particle creation by black holes}", \href{https://doi.org/10.1007/BF02345020}{ Commun.Math. Phys. 43 (1975) 199}.
\bibitem{Hawking3}
S. W. Hawking, ``\textit{Black holes and thermodynamics}", \href{ https://doi.org/10.1103/PhysRevD.13.191}{Phys. Rev. D 13 (1976) 191}.
\bibitem{Bekenstein1}
J. D. Bekenstein, ``\textit{Black Holes and the Second Law}", \href{ https://doi.org/10.1007/BF02762768}{Lett. Nuovo Cimento 4 (1972) 737}.
\bibitem{Bekenstein2}
J. D. Bekenstein, ``\textit{Black Holes and Entropy}", \href{ https://doi.org/10.1103/PhysRevD.7.2333}{Phys. Rev. D 7 (1973) 2333}.
\bibitem{Weiss}
P. Weiss, ``\textit{Black hole recipe: Slow light, swirl atoms}", \href{https://doi.org/10.2307/4012238}{Sci. News 157 (2000) 86}.
\bibitem{Philbin}
T. G. Philbin, C. Kuklewicz, S. Robertson, S. Hill, F. König,
and U. Leonhardt, ``\textit{Fiber-Optical Analog of the Event Horizon}", \href{10.1126/science.1153625}{Science 319 (2008) 1367}.
\bibitem{qsch}
S. Sen, R. Mandal and S. Gangopadhyay, ``\textit{Equivalence principle and HBAR entropy of an atom falling into a quantum corrected black hole}", \href{ https://doi.org/10.1103/PhysRevD.105.085007}{Phys. Rev. D 105 (2022) 085007}.
\bibitem{Camblong}
H. E. Camblong, A. Chakraborty and  C. R. Ord\'{o}\~{n}ez, ``\textit{Near-horizon aspects of acceleration radiation by free fall of an atom into a black hole}", \href{ https://doi.org/10.1103/PhysRevD.102.085010}{Phys. Rev. D 102 (2020) 085010}.
\bibitem{kerr1}
A. Azizi, H. E. Camblong, A. Chakraborty, C. R. Ord\'{o}\~{n}ez, and M. O. Scully,  ``\textit{Acceleration radiation of an atom freely falling into a Kerr black hole and near-horizon conformal quantum mechanics}", \href{https://link.aps.org/doi/10.1103/PhysRevD.104.065006}{Phys. Rev. D 104 (2021) 065006}.
\bibitem{AziziCamblong1}
A. Azizi,  H. E. Camblong, A. Chakraborty, C. R. Ord\'{o}\~{n}ez, and M. O. Scully, ``\textit{Quantum optics meets black hole thermodynamics via conformal quantum mechanics. I. Master equation for acceleration radiation}", \href{https://doi.org/10.1103/PhysRevD.104.084086}{Phys. Rev. D 104 (2021) 084086}.
\bibitem{AziziCamblong2}
A. Azizi,  H. E. Camblong, A. Chakraborty, C. R. Ord\'{o}\~{n}ez, and M. O. Scully, ``\textit{Quantum optics meets black hole thermodynamics via conformal quantum mechanics: II. Thermodynamics of acceleration radiation}", \href{ https://doi.org/10.1103/PhysRevD.104.084085}{Phys. Rev. D 104 (2021) 084085}.
\bibitem{SenMandal}
S. Sen, R. Mandal, and S. Gangopadhyay, ``\textit{Near horizon aspects  of acceleration radiation of an atom falling into a large class of static spherically symmetric black hole geometries}", \href{https://doi.org/10.1103/PhysRevD.106.025004}{Phys. Rev. D 106 (2022) 02500}.
\bibitem{kerr2}
S. Sen, R. Mandal, and S. Gangopadhyay, ``\textit{Near horizon approximation and beyond for a two-level atom falling into a Kerr–Newman black hole}", \href{https://doi.org/10.1140/epjp/s13360-023-04482-4}{Eur. Phys. J. Plus 138 (2023) 855}.
\bibitem{DasSen}
A. Das, S. Sen and S. Gangopadhyay, ``\textit{Horizon brightened accelerated radiation in the background of braneworld black holes}", \href{https://doi.org/10.1103/PhysRevD.109.064087}{Phys. Rev. D 109 (2024) 064087}.
\bibitem{KaulMajumdar}
R. K. Kaul and P. Majumdar, ``\textit{Logarithmic correction to the Bekenstein-Hawking entropy}", \href{https://doi.org/10.1103/PhysRevLett.84.5255}{Phys. Rev. Lett. 84 (2000) 5255}.
\bibitem{BonannoEichhorn}
 A. Bonanno, A. Eichhorn, H. Gies, J. M. Pawlowski, R. Percacci, M. Reuter, F. Saueressig, and G. P. Vacca, ``\textit{Critical Reflections on Asymptotically Safe Gravity}", \href{https://doi.org/10.3389/fphy.2020.00269 }{Front. Phys. 8 (2020) 269}.
\bibitem{hypergeometric}
\textit{Handbook of Mathematical Functions}, edited by M. Abramowitz and I. A. Stegun, (Dover, New York, 1965).
\bibitem{Boulware}
D. G. Boulware, ``\textit{Quantum field theory in Schwarzschild and Rindler spaces}", \href{ https://doi.org/10.1103/PhysRevD.11.1404}{Phys. Rev. D 11 (1975) 1404}.
\end{thebibliography}
\end{document}